\documentclass[runningheads]{llncs}
\usepackage{graphicx}
\usepackage{hyperref}
\usepackage[T1]{fontenc}
\usepackage{booktabs}
\usepackage{amsmath}
\usepackage[inline]{enumitem}
\usepackage{xspace}
\usepackage{caption}
\usepackage{subcaption}
\usepackage{xcolor}

\newcommand{\elbert}{ELBERT\xspace}

\begin{document}
\title{User Engagement Prediction for \\ Clarification in Search}
%
%
\author{Ivan Sekuli\'c\inst{1} \and
Mohammad Aliannejadi\inst{2} \and
Fabio Crestani\inst{1}}
\authorrunning{I. Sekuli\'c et al.}
%
\institute{Faculty of Informatics, 
Università della Svizzera italiana, 
Lugano, Switzerland \\
\email{\{ivan.sekulic,fabio.crestani\}@usi.ch},\\
\and
University of Amsterdam,
Amsterdam, The Netherlands \\
\email{m.aliannejadi@uva.nl}}
\maketitle              
\begin{abstract}

Clarification is increasingly becoming a vital factor in various topics of information retrieval, such as conversational search and modern Web search engines. Prompting the user for clarification in a search session can be very beneficial to the system as the user's explicit feedback helps the system improve retrieval massively. However, it comes with a very high risk of frustrating the user in case the system fails in asking decent clarifying questions. Therefore, it is of great importance to determine \textit{when} and \textit{how} to ask for clarification.

To this aim, in this work, we model search clarification prediction as user engagement problem. We assume that the better a clarification is, the higher user engagement with it would be. We propose a Transformer-based model to tackle the task. The comparison with competitive baselines on large-scale real-life clarification engagement data proves the effectiveness of our model. Also, we analyse the effect of all result page elements on the performance and find that, among others, the ranked list of the search engine leads to considerable improvements. Our extensive analysis of task-specific features guides future research.



\end{abstract}

\section{Introduction}

The primary goal of an information retrieval (IR) system is satisfying the user information need, which can often be ambiguous when expressed as short queries. 
Incorporating users' implicit feedback has long been studied for improved retrieval \cite{kelly2003implicit}. However, the recent rise of interest in conversational systems and mixed-initiative interactions have enabled IR systems to collect users' explicit feedback. Current research focuses on prompting users for feedback by asking for clarification~\cite{radlinski2017theoretical,aliannejadi2019asking,zamani2020generating}. 
For example, search clarification has recently been utilised in search engines, leading to an improved user experience \cite{zamani2020generating}.
Another prominent area studying clarification is conversational search, as the system can usually output only one response, thus requiring to clarify the user's intent~\cite{radlinski2017theoretical,aliannejadi2019asking}.
The importance of clarification further increases in a mixed-initiative conversational setting \cite{walker1990mixed}, where control of the conversation goes back and forth between user and system through assertions, prompts, and questions \cite{radlinski2017theoretical}.

However, clarification in search proved to be a cumbersome task \cite{zamani2020analyzing}, posing higher risk of user dissatisfaction.
The challenge arises from two main aspects: deciding whether or not it is necessary to ask for clarification, and selecting or generating the appropriate clarifying question.
Clarification selection can in fact be formalised as a user engagement prediction problem.
User engagement refers to the quality of user experience characterised  by, among others,  attributes  of positive affect,  attention, interactivity,  and  perceived  user  control \cite{o2008user}.
Persistent users' interactions with the clarification mechanism are an indication of a well-designed system.
Furthermore, through these interactions users provide implicit feedback about the \textit{necessity} and the \textit{quality} of prompted clarifications.

\begin{figure}[t]
    \centering
    \vspace{-4mm}
    \includegraphics[scale=0.25]{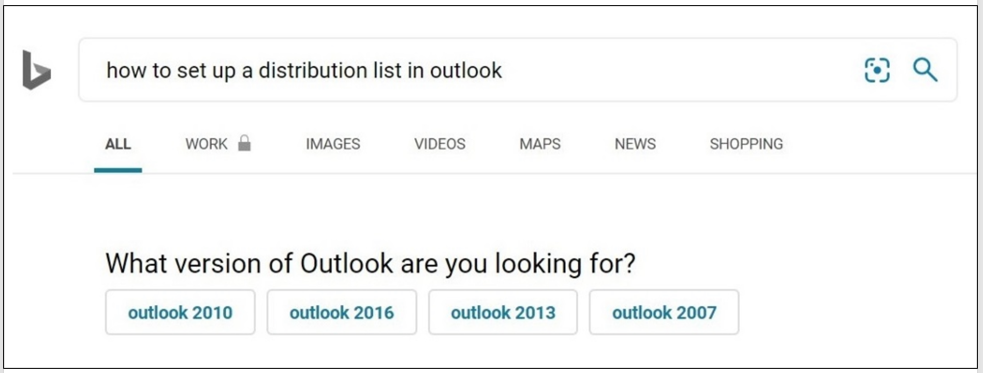}
    \caption{An example of Bing clarification pane taken from \cite{zamani2020mimics}.}
    \label{fig:my_label}
    \vspace{-4mm}
\end{figure}


Recently, modern search engines include various types of clarification components into their systems.
An example of such a component in Bing, namely a clarification pane, can be seen on Figure \ref{fig:my_label}.
Given a user query, a number of Microsoft's internal algorithms propose a clarifying question and offer clickable answers that would filter the retrieved results according to the user's need.
The research on the quality of asked clarifying questions and potential answers is still in its early stages \cite{zamani2020generating}; however, Zamani et al.~\cite{zamani2020mimics} argued that engagement level could be an indicator of the clarification system quality.
User engagement prediction has been studied in various domains of IR \cite{o2020empirical}. However, studying and modelling user engagement for web search clarification is relatively unstudied.


In this paper, we focus on the task of predicting user engagement level (ELP) on clarification panes.
Given an initial query, search results, and clarification pane, ELP aims to estimate how engaged the user would be with the clarification pane.
Previous work~\cite{zamani2020analyzing} studies how engagement levels correlate with the query attributes such as query type and aspects.
However, the relationship between SERPs and engagement has not yet been explored.
We stress the importance of utilising retrieved results, as they can contain cues as to how faceted or ambiguous the query is, suggesting how necessary the clarification is in the first place.

Moreover, users' engagement with the system implicitly discloses information about the \textit{necessity} and the \textit{quality} of the asked clarification.
The \textit{quality} aspect can be modelled under the assumption that the higher the engagement levels, the better the question and the provided answers are.
We make this assumption inspired by a large body of work in the IR community on implicit feedback from aggregated click-through rates for document retrieval \cite{xue2004optimizing}.
Also, we study clarification \textit{necessity} prediction through ELP. Our clarification necessity prediction model takes as input the initial query and the retrieved results list and predicts the level of user engagement with a clarification pane. Although certain attributes of the initial query such as length and ambiguity could indicate the necessity of asking clarifying questions, we show that incorporating other SERP elements such as result titles and snippets play important roles in improved prediction accuracy.

We formulate the task as supervised regression and propose a deep learning-based model for the prediction of the engagement levels.
We compare the performance of the model to various central tendency measures and a number of traditional machine learning algorithms, as well as popular neural models.
Our model, based on a Transformer architecture, jointly encodes the user query, the clarification pane, and the SERP elements, outperforming competitive baselines.
We evaluate the performance of our model on a large-scale dataset of search clarification engagements called MIMICS\footnote{\url{https://github.com/microsoft/MIMICS}}~\cite{zamani2020mimics}, collected from millions of interaction records of Bing\footnote{\url{http://www.bing.com}} users.
Our extensive experiments establish a strong baseline for the task, while ablation studies and analysis of the model's inner mechanisms provide guidelines for future research.
Our main contributions can be summarised as follows:
\begin{itemize}[label=\textbullet]
    \item We formally introduce the clarification pane ELP task as supervised regression and propose a transformer-based model to tackle it. We make the code publicly available for reproducibility purposes.\footnote{\url{https://github.com/isekulic/mimics-EL-benchmark}}.
    \item We perform ablation studies with respect to the model input data. We find that utilising retrieved search results greatly benefits the model's performance.
    \item We perform detailed analysis of the performance of our model w.r.t.~various characteristics of the SERP.
\end{itemize}

To the best of our knowledge, our work is the first to utilise SERP elements for clarification pane engagement prediction.
More precisely, we find that utilising search results in certain ways is highly beneficial for the ELP task, as the performance of our model increases by up to 40\% when provided with retrieved results, compared to the query and the clarification pane only.





\section{Related Work}
\label{rw}
Our work is related to work done in conversational and web search clarification, engagement level prediction, and neural networks. In this section we briefly review some of the works in these areas.

\textbf{Clarification.}
Search clarification has recently been addressed as an important problem in the IR community.
Recent research efforts study clarification in a wide range of areas, including 
web search engines \cite{zamani2020analyzing}, community question answering \cite{braslavski2017you}, voice queries \cite{kiesel2018toward}, dialogue systems \cite{stoyanchev2014towards}, entity disambiguation \cite{coden2015did}, and information-seeking conversations~\cite{aliannejadi2019asking,krasakis2020analysing,sekulic2020extending}.

Radlinski and Craswell \cite{radlinski2017theoretical} discuss the need for clarification in their proposed theoretical framework for conversational search, highlighting the necessity of multi-turn interactions with users.
Moreover, the report from the Dagstuhl Seminar on Conversational Search \cite{anand2020conversational} summarises potential research topics in conversational search, and recognises clarification as an integral part of a conversational information seeking (CIS) system, which was also argued by Penha et al.~\cite{penha2019introducing} for information-need elucidation.
Asking clarifying questions was studied by Aliannejadi et al.~\cite{aliannejadi2019asking}, who propose an offline evaluation setting of an open-domain CIS system, which was highlighted as a hard-to-evaluate setting \cite{penha2020challenges}.
They find that asking clarifying questions reduces the number of turns needed for identifying the underlying user information need.
Adding the fact that users like to be prompted for clarification \cite{kiesel2018toward}, we see a clear importance for clarification.

Clarification is further highlighted in mixed-initiative conversational search, where system in each turn needs to decide whether to ask for clarification or issue a response \cite{radlinski2017theoretical}.
Hashemi et al.~\cite{hashemi2020guided} propose a Guided Transformer model for document retrieval and next clarifying question selection in a conversational search setting.
Zamani et al.~\cite{zamani2020generating} propose supervised and reinforcement learning models for generating clarifying questions and the corresponding candidate answers from weak supervision data.
On the other hand, Ren et al.~\cite{ren2020conversations} introduce the task of conversations with search engines, where system generates a short, summarised response of the retrieved passages.
Although generating and selecting clarifying questions for such purposes has recently been studied, the necessity of asking for clarification is still a relatively unexplored topic~\cite{aliannejadi2020convai3}.
Whether or not it is necessary to ask for clarification depends mostly on the level of ambiguity of the query.

\textbf{User engagement.}
O'Brien and Toms \cite{o2008user} define user engagement as the quality of user experience in interaction with a system, characterised by various attributes, e.g., positive affect, aesthetic and sensory appeal, attention, novelty, perceived user control.
In their recent study \cite{o2020empirical}, they point user engagement as an important outcome measure in interactive IR research.
User engagement has previously been studied in the context of commercial software, social media \cite{di2016social}, online news \cite{o2017antecedents}, student engagement with online courses \cite{dhall2018emotiw}, and applications for monitoring health-related signals \cite{alkhaldi2016effectiveness}.




User engagement in the aforementioned studies has usually been measured by self-reported questionnaires, facial expression analysis or speech analysis, signal processing methods, or web analytics \cite{lalmas2014measuring}.
Recently, Zamani et al.~\cite{zamani2020mimics} created a collection of datasets for studying clarification in search by aggregating user interactions with clarification pane in a major commercial search engine, thus falling into the category of measuring the user engagement by web analytics.
In this paper however, instead of estimating the engagement levels with a goal of advancing search engine clarification feature, we analyse the implicit signals of the interactions that contain valuable information about the ambiguity of the query, diversity of retrieved results, and the quality of the clarifying question. 
Thus, motivated by work on implicit feedback of aggregated users' click-through logs for ad hoc retrieval \cite{kelly2003implicit},
we view the engagement levels as implicit evaluation of clarifying questions with respect to the query and search results.
Intuitively, the higher the engagement levels with the clarification system, the higher the quality of the prompted clarification, and higher the need for asking for clarification.


Zamani et al. \cite{zamani2020analyzing} study the clarifying question selection with respect to user queries, prompted questions and candidate answers in clarification panes of a search engine.
However, the retrieved search engine results for a query have not yet been studied.
To bridge this gap, in this paper, we propose a model to predict the user engagement levels, not only from the information in clarification pane, but from the retrieved search results. 

\textbf{Transformers.}
The unprecedented success of the Transformer-based architectures in the large variety of the IR and natural language processing tasks motivated their application to the engagement level prediction task as well. 
One of the most prominent Transformer-based models is BERT \cite{devlin2018bert}.
BERT has reached state-of-the-art results in multiple language understanding benchmarks, such as GLUE \cite{wang2018glue} and SQuAD \cite{rajpurkar2016squad}, as well as IR tasks, such as passage and document ranking \cite{nogueira2019passage,sekulic2020longformer}.
In this work, we utilise ALBERT \cite{lan2019albert} -- a lite BERT.
ALBERT offers the performance of BERT, or even a higher one, while having fewer parameters, reducing the GPU/TPU memory requirements.







\section{Engagement Level Prediction}
\label{method}
In this section, we first describe the dataset used for engagement level prediction (ELP).
Then, we formally introduce the task of ELP and propose a BERT-based model to tackle it.

\vspace{-4mm}
\subsection{Data}
\label{dataformat}

MIMICS \cite{zamani2020mimics} is a recently proposed large-scale collection of datasets for research on search clarification.
It enables the IR community to study various aspects of search clarification, ranging from clarification generation and selection,  over re-ranking of candidate answers, to user engagement prediction and click models for clarification. MIMICS consists of three datasets:

\begin{enumerate}
    \item \textbf{MIMICS-Click}, including over 400k unique queries, their corresponding clarification panes, and the aggregated user interaction signals.
    \item \textbf{MIMICS-ClickExplore}, consisting of over 60k unique queries, each with multiple clarification panes, and the aggregated interaction signals.
    \item \textbf{MIMICS-Manual}, containing 2k query-clarification pairs, manually labelled for the quality of clarifying questions, candidate answer sets, and landing result pages of each answer.
\end{enumerate}

In this work, we mainly focus on MIMICS-Click, as the largest, most generic one.
Each sample in MIMICS-Click consists of the initial query $q$, the clarification question $c$, and answers offered as options by the system $A = [a_1,...a_5]$.
The sample is associated with user interaction signals as labels.
The \textit{impression level}~$i$, a categorical variable where $i \in \{low,medium,high\}$, represents the frequency of the clarification pane being presented to the user for the corresponding query.
The \textit{engagement level} $e \in [0, 10]$ shows the level of total engagement received by the users in terms of click-through rate.
Each answer is also associated with its conditional click probability.

The authors also released search engine results pages (SERPs) for each query, as retrieved by Bing.
In addition to the query meta-data, SERPs contain up to $10$ retrieved instances with a title, an URL, and a short snippet of a web document.
We denote retrieved results as $R = [r_1,r_2,\dots r_n]$, where $n \in [0,10]$. Each of the results $r_i$ consists of a tuple $r_i = (t_i, s_i)$, where $t_i$ and $s_i$ are title and snippet of the $i$-th result.
Table \ref{tbl:stats} shows the average lengths of queries\footnote{The length was computed by splitting the text on whitespaces.}, questions, retrieved titles and snippets, as well as the number of retrieved results in SERPs.
We utilise all of the available text and information as input to our models to compose our experiments, as described in Section \ref{models}.


\begin{table}[t]
\vspace{-5mm}
\centering
\caption{Dataset statistics for MIMICS-Click.}
\begin{tabular}{l@{\quad}l@{\quad}l@{\quad}l@{\quad}l}
\toprule
 & Mean & Std & Median & min-max \\
 \midrule
Query length       & 2.66   & 1.18  & 2    & 1 - 12       \\
Question length    & 6.05   & 0.47   & 6     & 5 - 14        \\
SERP Titles length      & 7.65   & 2.71   & 8     & 0 - 30        \\
SERP Snippets length    & 43.47   & 14.76   & 45     & 0 - 149          \\
Answers per query      & 2.81 & 1.06 & 2 & 2 - 5 \\
Responses per query       & 9.07   & 1.19     & 9    & 0 - 10 \\
\bottomrule 
\end{tabular}
\label{tbl:stats}
\vspace{-5mm}
\end{table}


\vspace{-2mm}
\subsection{Task Formulation}
We formulate the task of user engagement level prediction as a supervised regression.
The goal of the regression is to predict the value of the target variable $y$, given a D-dimensional vector $\textbf{x}$ of input variables \cite{bishop2006pattern}.
Given the dataset of $N$ observation pairs $(\textbf{x}_n, y_n)$, where $n = 1,...,N$, the goal is to find a function $f(\textbf{x})$ whose outputs $\hat{y}$ for new inputs $\textbf{x}$ produce the predictions for the corresponding values of $y$.
The loss function of the predicted values $\hat{y}$ and the actual values $y$ are model-dependent and described in Section \ref{models}.


The target variable $y$ is given in the dataset in the range of 0 to 10, corresponding to the level of user engagement with the clarification pane. 
We approach ELP as a regression problem as it poses itself as a natural formulation of our task. Compared to classification, false predictions of different value are penalised differently. For example, classification would punish false predictions of $\hat{y}=7$ and $\hat{y}=1$ for a sample with $y=8$ the same, while in reality, the predicted label of $7$ is much closer to the actual engagement level. Therefore, even though still wrong, one would prefer a system to predict 7 instead of 1. Moreover, the task of user engagement prediction has been evaluated as regression in various applications such as \cite{sano2016prediction,dhall2018emotiw}.

\subsection{Our approach}
\label{models}
We now define our model called \textbf{ELBERT} (\textbf{E}ngagement \textbf{L}evel prediction by A\textbf{LBERT}).
As mentioned in the previous section, the goal is to predict the engagement level $y$ based on the initial query $q$, clarification question $c$, list of candidate answers $A$, and retrieved results $R$. 
We predict the engagement level $EL$ as follows:
\begin{equation}
\label{eq:el}
    EL(q,c,A,R) = \psi(\phi_q(q),\phi_c(c),\phi_A(A),\phi_R(R))
\end{equation}
where $\phi_{\{q,c,A,R\}}$ are high-dimensional representations of $q$, $c$, $A$, and $R$. The aggregation function $\psi$ outputs the final engagement levels based on the input representations.
All of these components can be modelled with numerous methods.
In this work, we utilise ALBERT as our encoder for generating $\phi_{\{q,c,A,R\}}$ representations in a joint fashion.
More specifically, as ALBERT has been shown to consistently help downstream tasks with multiple inputs \cite{lan2019albert}, we essentially learn the joint representation of query, clarification question, answers, and results as:
\begin{equation}
\label{eq:albert}
    \Phi(q,c,A,R) = ALBERT(q,c,A,R)
\end{equation}
reducing our Equation \ref{eq:el} to:
\begin{equation}
    EL(q,c,A,R) = \psi(\Phi(q,c,A,R)).
\end{equation}
Input to the ALBERT component is composed of tokenized query, question, answers, and results, separated by the separation token $[SEP]$, with classification token $[CLS]$ inserted in the beginning of a sequence. 
Answers $a_i$ are aggregated before feeding them to the model.
Similarly, we aggregate SERP information $R$, with a difference that we experiment with both, titles $t_i$ and snippets $s_i$ as inputs. 
In either case, texts of titles or of snippets are joined by whitespace prior to being fed to the model.
We note that in ablation studies some of the components are left out by simply removing them from Equation \ref{eq:albert}.
We use a pretrained ALBERT-base \cite{lan2019albert} as a text encoder and truncate the total input sequence length to a maximum of $512$ tokens.
Our model has $11M$ training parameters, making it considerably smaller than other Transformer-based model (e.g., BERT has $110M$).

The regression component $\psi$, that outputs the engagement level, is constructed as follows:
last layer hidden-state of the first token of the encoded sequence ($[CLS]$ token) is further processed by a linear layer and a non-linear activation function. 
We then add another linear layer, with dropout and a non-linear activation function in between, to produce the final 1-dimensional output that corresponds to $EL$.
The model is trained using mean squared error as a loss function for 4 epochs, with a learning rate of $5\times10^{-5}$, Adam optimizer \cite{kingma2014adam} and linear weight decay with warmup.


\section{Experiments}
\label{experiments}
In this section, we introduce our experimental setup and present main results for the engagement level prediction.
Furthermore, we analyse the effect of SERP elements on model's performance and perform detailed analysis w.r.t.~various characteristics of the data.


\subsection{Baselines}
We use central tendency measures as our first baselines for predicting the engagement level. More specifically, we have three different static baselines:
\begin{enumerate*}[label=(\roman*)]
    \item \textit{mean} of the data (MeanEngagement);
    \item \textit{median} of the data (MedianEngagement);
    \item \textit{sampling} from a normal distribution $\mathcal{N}(\mu, \sigma^{2})$, where $\mu$ and $\sigma$ are the mean and the standard deviation of the engagement levels in the training data, respectively (NormalEngagement).
\end{enumerate*}

To tackle the task of ELP, we experiment with a number of models from traditional machine learning and deep learning. Namely:
\begin{description}
    \item[Linear Regression.] First baseline is a linear regression model, fitted using ordinary least squares approach.
    \item[SVR.] We employ support vector regression machines \cite{drucker1997support}, a version of support vector machines \cite{cortes1995support} for regression. We experiment with the linear, as well as the radial basis function (RBF) kernel.
    \item[Random Forests.] An ensemble meta-algorithm that uses bootstrap aggregating (bagging) technique to improve the stability of decision trees \cite{breiman2001random}. 
    \item[LSTM.] Long-short term memory \cite{hochreiter1997long} are a well-established method for sequence modelling, especially on text data. We experiment with multi-layer bidirectional networks.
\end{description}
The input to traditional ML models are tf-idf weighted bag-of-word features extracted from the input text.
LSTM is fed with pretrained GloVe word embeddings \cite{pennington2014glove} of tokenized input text.
We use Scikit-learn \cite{pedregosa2011scikit}, HuggingFace \cite{wolf2019huggingface}, and Pytorch \cite{paszke2019pytorch} for the implementation of the aforementioned models.

\subsection{Evaluation Metrics}
We evaluate the effectiveness of our models using standard evaluation metrics for the task of supervised regression. 
The first two are Mean Absolute Error (MAE) and Mean Squared Error (MSE).
We also evaluate our regression models with Coefficient of Determination or $R^{2}$.
It is a statistical measurement that examines the proportion of the variance in one variable that is predictable from the second variable, estimating the \textquotedblleft goodness of a fit\textquotedblright. It is defined as: $
R^2 = 1 - \frac{\sum_{i=1}^N (y_i - \hat{y}_i)^2}{\sum_{i=1}^N (y_i - \overline{y}_i)^2}$,
where $N$ is the number of samples, $y_i$ is the actual value in the dataset for the $i$-th sample, $\hat{y}$ is the predicted value, and $\overline{y}$ is the mean of the actual values. 

\subsection{Experimental Setup}
We evaluate our models using a hold-out method, i.e., reserving 20\% of the dataset for the test set. 
We train, and tune traditional ML models in a cross-validation manner \cite{cawley2010over}.
We use 5-fold split of the training set into training and development set, which is used for grid-searching of the best parameters.
The extensive grids of parameters include regularisation parameter \textit{C}, the choice of \textit{kernel}, \textit{gamma}, and \textit{epsilon} for SVR, number of estimators and depth of random forest regressor, as well as feature selection process. 
All of the parameters can be found on our GitHub repository.

For tuning the hyper-parameters of our neural models, we split the training set into training and development sets.
Notice that models are retrained on the full training set with the best parameters before being evaluated on the hold-out test set.

We evaluate the models on the full MIMICS-Click dataset, consisting of more than 400k query-clarification-SERP tuples, and on the subset of that dataset, in which only samples with the engagement level larger than zero are selected. 
The models in this setting were fed all the available data, i.e., the queries, clarification panes, and the SERPs, while the ablation studies in Section~\ref{ablation} go into the analysis of input data. 


\subsection{Results \& Discussion}

\subsubsection{Performance comparison.}

\begin{table}[t]
\centering
\vspace{-4mm}
\caption{Performance on the full MIMICS-Click dataset (400k+ samples) and a subset where engagement levels are higher than zero (71k samples). Bold values denote the best results for each metric. Symbols $\dagger$ and $\ddag$ mark statistically significant improvement over central tendency measures and traditional ML models, respectively ($p < 0.01$).}
\begin{tabular}{l@{\quad}l@{\quad}ll@{\quad}l@{\quad}l@{\quad}l@{\quad}l}
 \toprule
                      & \multicolumn{3}{c}{Full MIMICS-Click}       &        & \multicolumn{3}{c}{EL-only MIMICS-Click} \\
                      \cmidrule{2-4}  \cmidrule{6-8}
Model                 & MAE             & MSE             & $R^2$       &    & MAE      & MSE      & $R^2$     \\
\midrule
Mean                  & 0.1531          & 0.0546          & 0.0       &      & 0.2426  & 0.0790   &  0.0  \\
Median                & \textbf{0.0921}$^{\dagger}$ & 0.0531         & 0.0      &       & 0.2412 &  0.0805  &  0.0    \\
Normal                & 0.1896          & 0.0823          & 0.0       &      & 0.4316   & 0.2637   &  0.0 \\
\midrule
Linear Regression     & 0.1463          & 0.0530          & 0.0359     &     & 0.2364   &  0.0783  &  0.0083      \\
SVR                   & 0.1462          & 0.0522          & 0.0490      &     & $0.2318^{\dagger}$  &  $0.0736^{\dagger}$  & $0.0676^{\dagger}$  \\
RandomForest          & 0.1477          & 0.0526          & 0.0423  &&  $0.2301^{\dagger}$  &  $0.0729^{\dagger}$  &  $0.0775^{\dagger}$   \\
\midrule
BiLSTM                & $0.1452^{\dagger\ddag}$ & $0.0511^{\dagger\ddag}$ & $0.0606^{\dagger\ddag}$ && $0.2299^{\dagger}$ & $0.0720^{\dagger}$ & $0.0789^{\dagger}$ \\
ELBERT                & $0.1439^{\dagger\ddag}$ & \textbf{0.0505}$^{\dagger\ddag}$ & \textbf{0.0762}$^{\dagger\ddag}$        & & \textbf{0.2224}$^{\dagger\ddag}$   &  \textbf{0.0692}$^{\dagger\ddag}$  &  \textbf{0.1124}$^{\dagger\ddag}$ \\
\bottomrule
\end{tabular}
\vspace{-4mm}
\label{tbl:results}
\end{table}

Here, we compare the performance of our \elbert model against the baselines on the complete dataset, as well as the subset of data with EL > 0. Table~\ref{tbl:results} lists the results in terms of all our evaluation metrics. 
We can notice that heuristic baselines (i.e., MeanEngagement, MedianEngagement and NormalEngagement) are consistently outperformed by both, the traditional ML models, and the neural models.
However, one exception is MedianEngagement, a baseline that always outputs the median of the training set, i.e., EL of 0.0, when evaluated on the full MIMICS-Click by mean absolute error.
Since more than $80\%$ of the dataset have EL of 0.0, and MAE does not penalise large errors as hard as MSE or $R^2$, this is expected. 
The tide turns swiftly when evaluating on the subset of the data with EL larger that $0.0$, where all of the static baselines, including MedianEngagement, are outperformed by all of our models.

Moreover, we see a clear disparency in the performance of traditional ML models and neural networks.
This is consistent with recent research in various tasks in IR and NLP fields.
Moreover, we see that \elbert significantly outperforms BiLSTM model.
Through its powerful encoder, ELBERT is able to capture deeper semantic relations, as it is pretrained on a large body of text.
This is also consistent with recent research on deep learning-based models for natural language understanding.


\vspace{-4mm}
\subsubsection{Effect of SERP elements on ELP.}
\label{ablation}


\begin{table}[t]
\vspace{-3mm}
\centering
\caption{Impact of SERP elements available on the model performance. Bold values denote the best performance of each metric. Statistically significant results (with $p < 0.05$) over \textit{query} setting and \textit{query+pane} setting are marked with ${\dagger}$ and ${\ddag}$, respectively.}
\resizebox{\textwidth}{!}{
\begin{tabular}{ll@{\quad}l@{\quad}ll@{\quad}l@{\quad}l@{\quad}l@{\quad}l}
 \toprule
   &                   & \multicolumn{3}{c}{Full MIMICS-Click}       &        & \multicolumn{3}{c}{EL-only MIMICS-Click} \\
                      \cmidrule{3-5}  \cmidrule{7-9}
\# & Setting                 & MAE             & MSE             & $R^2$           && MAE      & MSE      & $R^2$     \\
\midrule
1 & query                 & 0.1500 &  0.0519 & 0.0485 && 0.2275 & 0.0719 & 0.0776 \\
2 & query+pane            & 0.1354$^{\dagger}$ & 0.0512 & 0.0626$^{\dagger}$ && 0.2257$^{\dagger}$ & 0.0714 & 0.0839$^{\dagger}$ \\
3 & query+titles          & \textbf{0.1335}$^{\dagger\ddag}$ & \textbf{0.0436}$^{\dagger\ddag}$ & \textbf{0.0814}$^{\dagger\ddag}$ && 0.2229$^{\dagger\ddag}$ & \textbf{0.0692}$^{\dagger\ddag}$  & \textbf{0.1124}$^{\dagger\ddag}$ \\
4 & query+snippets        & 0.1459$^{\dagger}$ & 0.0513 & 0.0606$^{\dagger}$ && 0.2255$^{\dagger\ddag}$ & 0.0706$^{\dagger\ddag}$ & 0.0944$^{\dagger\ddag}$ \\
5 & query+pane+titles     & 0.1450$^{\dagger}$ & 0.0505$^{\dagger}$ & 0.0745$^{\dagger\ddag}$ && \textbf{0.2224}$^{\dagger\ddag}$ & \textbf{0.0692}$^{\dagger\ddag}$ & \textbf{0.1124}$^{\dagger\ddag}$ \\
6 & query+pane+snippets   & 0.1439$^{\dagger}$ & 0.0505$^{\dagger}$ & 0.0762$^{\dagger\ddag}$ && 0.2240$^{\dagger\ddag}$ & 0.0704$^{\dagger\ddag}$ & 0.0969$^{\dagger\ddag}$ \\
\bottomrule \\
\end{tabular}}
\label{tbl:ablation1}
\vspace{-3mm}
\end{table}


In this experiment, we aim to analyse the effect of clarification panes and every SERP element on the performance of our model. Our hypothesis is that each SERP element (e.g., result titles and snippets) provides a complementary set of features that aids the model towards more effective prediction. Therefore, we train our \elbert model with different combination of SERP elements and clarification panes, and compare the performance of the different models. We report the results in Table~\ref{tbl:ablation1}.
We see that the relative improvement when utilising titles from SERPs is up to $35\%$ compared to using query and clarification pane, and more than $45\%$ over query-only setting.
The results strongly suggest the advantage of making use of SERP elements for ELP.

An interesting finding is that even though snippets contain more text than titles and thus arguably more information as well, the model does not consistently perform better with snippets as input. 
In fact, even though results with titles seem better than ones with snippets, we observe no statistically significant difference between the performance of \textit{query+titles} and \textit{query+snippets} on full MIMICS-Click, nor EL-only MIMICS-Click.
There are several reasons why snippets do not exceed the performance of titles.
First, it might be the quality and type of text shown in snippets.
Snippets often show only short excerpts, or even multiple excerpts which are not clearly divided, from a longer document, focusing on query words in the retrieved document. 
Thus, they might not contain all the semantics of the document, while titles usually do.
Second, it might be the maximum input length of our encoder, which is 512 sub-word tokens.
As mentioned in Table \ref{tbl:stats}, a median length of a title is 8 tokens, while median snippet length is $45$.
Considering that most of the samples have 9 or more title-snippet pairs in their SERPs, it is evident that some portion of concatenated snippets get left out.
The potential limitation of truncating input length in most of BERT-based models is a research direction on its own.

We point out that the necessity of asking the clarification can be estimated from the initial query and retrieved search results, i.e., rows $1$, $3$, and $4$ in Table \ref{tbl:ablation1},
The success of the model to predict EL based on SERPs and the query alone, suggests that this framework can be used for determining whether or not to ask a clarifying question.
However, we leave this aspect for future work.
Instead, in the next subsection we evaluate our model trained on ELP task for clarification pane selection, addressing the pane quality aspect.


\vspace{-2mm}
\subsection{Additional Experiments}

Here we show \elbert performance, as measured by $R^2$, with respect to various characteristics of the dataset and the input components.

\vspace{-3mm}
\subsubsection{Impression level.}
\begin{figure}[t]
    \centering
    \includegraphics[width=\textwidth]{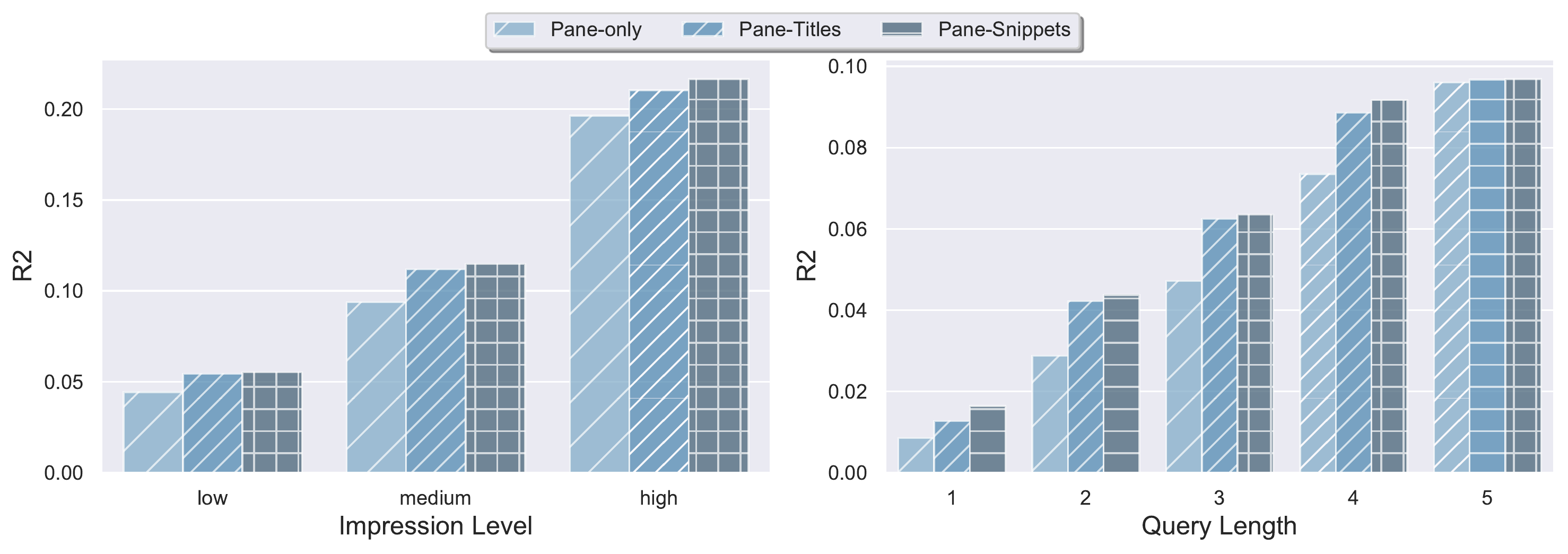}
    \caption{Performance by impression levels (left) and query lengths (right) with different input configurations.}
    \label{fig:final}
    \vspace{-3mm}
\end{figure}

Figure \ref{fig:final} (left) shows the performance of our model w.r.t.~impression levels.
We notice that our model performs significantly better on queries with high impression rate, i.e., those whose clarification panes have been shown to users more frequently.
The differences between models at each impression level are not statistically significant, while differences between levels are, with $p < 0.01$.
As the engagement level labels have been computed by aggregating user click information, this suggest that query-clarification pairs that have been implicitly evaluated by a small number of users, i.e., have low impression level, contain noise.

\vspace{-5mm}
\subsubsection{Query length.} 
Figure \ref{fig:final} (right) presents the performance of our model w.r.t.~query length.
The difference in performance between all query lengths is statistically significant.
We notice that longer queries generally lead to better performance.
This can be attributed to them being more descriptive, thus allowing the search engine to retrieve more relevant results.
Consequently, our model would utilise SERPs of higher quality, improving the ELP.
Highest improvement is seen for a query and pane-only setting.
Since the model in that setting does not see any SERP content, it benefits the most out of longer, more descriptive queries.


\vspace{-5mm}
\subsubsection{Number of search results.}
\begin{figure}[t]
\vspace{-4mm}
  \centering
  \includegraphics[scale=0.5]{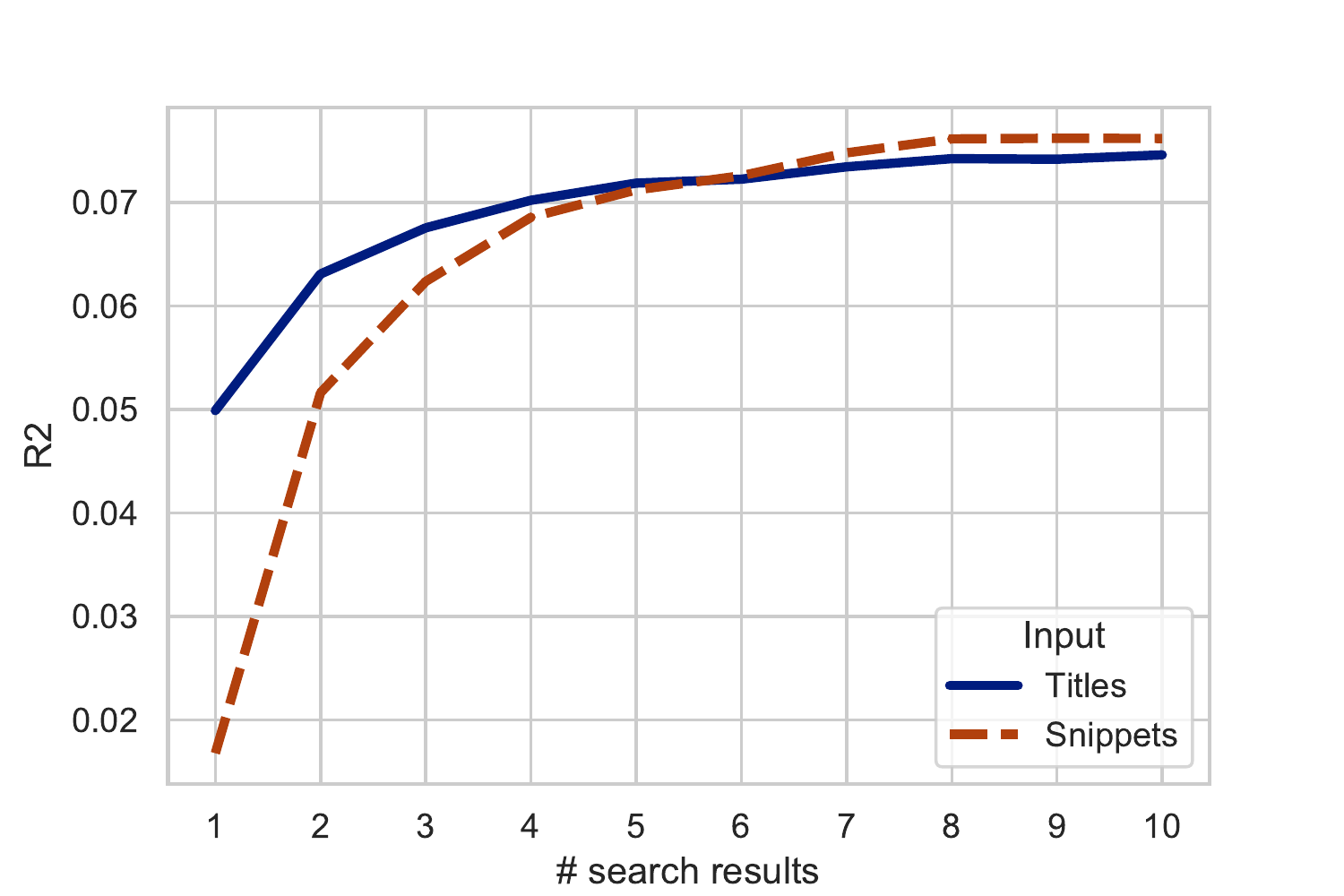}
  \vspace{-3mm}
  \caption{Performance by number of search results made available to the model.}
  \label{fig:nserp}
  \vspace{-2mm}
\end{figure}

Since user behaviour is mainly biased by the results they see, and they mostly look at top results only, we perform experiments to see how our models behave in a setting with limited number of retrieved results.
As mentioned before, MIMICS dataset contains up to $10$ retrieved results for each query.
We evaluate our model with $1,2,\dots10$ SERP elements made available to it.
Results for both, titles setting and snippets setting, are presented in Figure \ref{fig:nserp}.
We see a clear improvement in the performance as the number of search results fed to the model rises.
This suggests that our model highly utilises SERP elements for ELP. 
We notice a saturation after 7 elements, especially in the setting with snippets.
This might be due to snippets exceeding the maximum length of input to transformer-based models, which is 512 subword tokens.


\vspace{-2mm}
\section{Conclusions}
\label{conclusion}
In this study, we conducted various experiments on engagement level prediction task for clarification in search.
We showed that semantic-rich models, like ALBERT, are much more successful in the task than traditional ML models.
Furthermore, we demonstrated the benefit of utilising information from search engine result pages, such as titles and text snippets of retrieved documents, in the ELP task.
Modelling of engagement levels can help guide the system on when and which clarifications to prompt, thus improving the overall user experience.
Future work involves deeper analysis of topical changes in the retrieved pages, that could lead to more accurate prediction of engagement levels, and estimating the necessity of asking for clarification.


%
%
%
\bibliographystyle{splncs04}
\bibliography{mybib}

\begin{thebibliography}{10}
\providecommand{\url}[1]{\texttt{#1}}
\providecommand{\urlprefix}{URL }
\providecommand{\doi}[1]{https://doi.org/#1}

\bibitem{aliannejadi2020convai3}
Aliannejadi, M., Kiseleva, J., Chuklin, A., Dalton, J., Burtsev, M.: Conv{AI}3:
  Generating clarifying questions for open-domain dialogue systems ({ClariQ})
  (2020)

\bibitem{aliannejadi2019asking}
Aliannejadi, M., Zamani, H., Crestani, F., Croft, W.B.: Asking clarifying
  questions in open-domain information-seeking conversations. In: Proceedings
  of the 42nd International ACM SIGIR Conference on Research and Development in
  Information Retrieval. pp. 475--484 (2019)

\bibitem{alkhaldi2016effectiveness}
Alkhaldi, G., Hamilton, F.L., Lau, R., Webster, R., Michie, S., Murray, E.: The
  effectiveness of prompts to promote engagement with digital interventions: a
  systematic review. Journal of medical Internet research  \textbf{18}(1), ~e6
  (2016)

\bibitem{anand2020conversational}
Anand, A., Cavedon, L., Joho, H., Sanderson, M., Stein, B.: Conversational
  search ({D}agstuhl {S}eminar 19461). In: Dagstuhl Reports. vol.~9. Schloss
  Dagstuhl-Leibniz-Zentrum f{\"u}r Informatik (2020)

\bibitem{bishop2006pattern}
Bishop, C.M.: Pattern Recognition and Machine Learning. Springer (2006)

\bibitem{braslavski2017you}
Braslavski, P., Savenkov, D., Agichtein, E., Dubatovka, A.: What do you mean
  exactly? {A}nalyzing clarification questions in {CQA}. In: Proceedings of the
  2017 Conference on Conference Human Information Interaction and Retrieval.
  pp. 345--348 (2017)

\bibitem{breiman2001random}
Breiman, L.: Random forests. Machine learning  \textbf{45}(1),  5--32 (2001)

\bibitem{cawley2010over}
Cawley, G.C., Talbot, N.L.: On over-fitting in model selection and subsequent
  selection bias in performance evaluation. The Journal of Machine Learning
  Research  \textbf{11},  2079--2107 (2010)

\bibitem{coden2015did}
Coden, A., Gruhl, D., Lewis, N., Mendes, P.N.: Did you mean {A} or {B}?
  {S}upporting clarification dialog for entity disambiguation. In: SumPre-HSWI@
  ESWC (2015)

\bibitem{cortes1995support}
Cortes, C., Vapnik, V.: Support-vector networks. Machine learning
  \textbf{20}(3),  273--297 (1995)

\bibitem{devlin2018bert}
Devlin, J., Chang, M.W., Lee, K., Toutanova, K.: {BERT}: Pre-training of deep
  bidirectional transformers for language understanding. In: Proceedings of the
  2019 Conference of the North {A}merican Chapter of the Association for
  Computational Linguistics: Human Language Technologies, Volume 1 (Long and
  Short Papers). pp. 4171--4186. Association for Computational Linguistics,
  Minneapolis, Minnesota (Jun 2019). \doi{10.18653/v1/N19-1423},
  \url{https://www.aclweb.org/anthology/N19-1423}

\bibitem{dhall2018emotiw}
Dhall, A., Kaur, A., Goecke, R., Gedeon, T.: Emotiw 2018: Audio-video, student
  engagement and group-level affect prediction. In: Proceedings of the 20th ACM
  International Conference on Multimodal Interaction. pp. 653--656 (2018)

\bibitem{di2016social}
Di~Gangi, P.M., Wasko, M.M.: Social media engagement theory: Exploring the
  influence of user engagement on social media usage. Journal of Organizational
  and End User Computing (JOEUC)  \textbf{28}(2),  53--73 (2016)

\bibitem{drucker1997support}
Drucker, H., Burges, C.J., Kaufman, L., Smola, A.J., Vapnik, V.: Support vector
  regression machines. In: Advances in neural information processing systems.
  pp. 155--161 (1997)

\bibitem{hashemi2020guided}
Hashemi, H., Zamani, H., Croft, W.B.: Guided transformer: Leveraging multiple
  external sources for representation learning in conversational search. In:
  Proceedings of the 43rd International ACM SIGIR Conference on Research and
  Development in Information Retrieval. pp. 1131--1140 (2020)

\bibitem{hochreiter1997long}
Hochreiter, S., Schmidhuber, J.: Long short-term memory. Neural computation
  \textbf{9}(8),  1735--1780 (1997)

\bibitem{kelly2003implicit}
Kelly, D., Teevan, J.: Implicit feedback for inferring user preference: a
  bibliography. In: Acm Sigir Forum. vol.~37, pp. 18--28. ACM New York, NY, USA
  (2003)

\bibitem{kiesel2018toward}
Kiesel, J., Bahrami, A., Stein, B., Anand, A., Hagen, M.: Toward voice query
  clarification. In: The 41st International ACM SIGIR Conference on Research \&
  Development in Information Retrieval. pp. 1257--1260 (2018)

\bibitem{kingma2014adam}
Kingma, D.P., Ba, J.: Adam: A method for stochastic optimization. arXiv
  preprint arXiv:1412.6980  (2014)

\bibitem{krasakis2020analysing}
Krasakis, A.M., Aliannejadi, M., Voskarides, N., Kanoulas, E.: Analysing the
  effect of clarifying questions on document ranking in conversational search.
  In: Proceedings of the 2020 ACM SIGIR on International Conference on Theory
  of Information Retrieval. pp. 129--132 (2020)

\bibitem{lalmas2014measuring}
Lalmas, M., O'Brien, H., Yom-Tov, E.: Measuring user engagement. Synthesis
  Lectures on Information Concepts, Retrieval, and Services  \textbf{6}(4),
  1--132 (2014)

\bibitem{lan2019albert}
Lan, Z., Chen, M., Goodman, S., Gimpel, K., Sharma, P., Soricut, R.: Albert: A
  lite {BERT} for self-supervised learning of language representations. In
  proceedings of ICLR  (2020)

\bibitem{nogueira2019passage}
Nogueira, R., Cho, K.: Passage re-ranking with bert. arXiv preprint
  arXiv:1901.04085  (2019)

\bibitem{o2017antecedents}
O'Brien, H.L.: Antecedents and learning outcomes of online news engagement.
  Journal of the Association for Information Science and Technology
  \textbf{68}(12),  2809--2820 (2017)

\bibitem{o2020empirical}
O'Brien, H.L., Arguello, J., Capra, R.: An empirical study of interest, task
  complexity, and search behaviour on user engagement. Information Processing
  \& Management  \textbf{57}(3),  102226 (2020)

\bibitem{o2008user}
O'Brien, H.L., Toms, E.G.: What is user engagement? {A} conceptual framework
  for defining user engagement with technology. Journal of the American society
  for Information Science and Technology  \textbf{59}(6),  938--955 (2008)

\bibitem{paszke2019pytorch}
Paszke, A., Gross, S., Massa, F., Lerer, A., Bradbury, J., Chanan, G., Killeen,
  T., Lin, Z., Gimelshein, N., Antiga, L., et~al.: Pytorch: An imperative
  style, high-performance deep learning library. In: Advances in neural
  information processing systems. pp. 8026--8037 (2019)

\bibitem{pedregosa2011scikit}
Pedregosa, F., Varoquaux, G., Gramfort, A., Michel, V., Thirion, B., Grisel,
  O., Blondel, M., Prettenhofer, P., Weiss, R., Dubourg, V., et~al.:
  Scikit-learn: Machine learning in {P}ython. the Journal of machine Learning
  research  \textbf{12},  2825--2830 (2011)

\bibitem{penha2019introducing}
Penha, G., Balan, A., Hauff, C.: Introducing {MANtIS}: a novel multi-domain
  information seeking dialogues dataset. arXiv preprint arXiv:1912.04639
  (2019)

\bibitem{penha2020challenges}
Penha, G., Hauff, C.: Challenges in the evaluation of conversational search
  systems. KDD Workshop on Conversational Systems Towards Mainstream Adoption
  (2020)

\bibitem{pennington2014glove}
Pennington, J., Socher, R., Manning, C.D.: Glove: Global vectors for word
  representation. In: Proceedings of the 2014 conference on empirical methods
  in natural language processing (EMNLP). pp. 1532--1543 (2014)

\bibitem{radlinski2017theoretical}
Radlinski, F., Craswell, N.: A theoretical framework for conversational search.
  In: Proceedings of the 2017 conference on conference human information
  interaction and retrieval. pp. 117--126 (2017)

\bibitem{rajpurkar2016squad}
Rajpurkar, P., Zhang, J., Lopyrev, K., Liang, P.: {SQ}u{AD}: 100,000+ questions
  for machine comprehension of text. In: Proceedings of the 2016 Conference on
  Empirical Methods in Natural Language Processing. pp. 2383--2392. Association
  for Computational Linguistics, Austin, Texas (Nov 2016)

\bibitem{ren2020conversations}
REN, P., CHEN, Z., REN, Z., KANOULAS, E., MONZ, C., DE~RIJKE, M.: Conversations
  with search engines. ACM Transactions on Information Systems  \textbf{1}(1)
  (2020)

\bibitem{sano2016prediction}
Sano, S., Kaji, N., Sassano, M.: Prediction of prospective user engagement with
  intelligent assistants. In: Proceedings of the 54th Annual Meeting of the
  Association for Computational Linguistics (Volume 1: Long Papers). pp.
  1203--1212 (2016)

\bibitem{sekulic2020extending}
Sekuli{\'c}, I., Aliannejadi, M., Crestani, F.: Extending the use of previous
  relevant utterances for response ranking in conversational search. In:
  Proceedings of the Twenty-Ninth Text REtrieval Conference, {TREC} (2020)

\bibitem{sekulic2020longformer}
Sekuli{\'c}, I., Soleimani, A., Aliannejadi, M., Crestani, F.: Longformer for
  {MS} {MARCO} document re-ranking task. arXiv preprint arXiv:2009.09392
  (2020)

\bibitem{stoyanchev2014towards}
Stoyanchev, S., Liu, A., Hirschberg, J.: Towards natural clarification
  questions in dialogue systems. In: AISB symposium on questions, discourse and
  dialogue. vol.~20 (2014)

\bibitem{walker1990mixed}
Walker, M.A., Whittaker, S.: Mixed initiative in dialogue: An investigation
  into discourse segmentation. In: ACL (1990)

\bibitem{wang2018glue}
Wang, A., Singh, A., Michael, J., Hill, F., Levy, O., Bowman, S.: Glue: A
  multi-task benchmark and analysis platform for natural language
  understanding. In: Proceedings of the 2018 EMNLP Workshop BlackboxNLP:
  Analyzing and Interpreting Neural Networks for NLP. pp. 353--355 (2018)

\bibitem{wolf2019huggingface}
Wolf, T., Debut, L., Sanh, V., Chaumond, J., Delangue, C., Moi, A., Cistac, P.,
  Rault, T., Louf, R., Funtowicz, M., et~al.: Huggingface's transformers:
  State-of-the-art natural language processing. ArXiv pp. arXiv--1910 (2019)

\bibitem{xue2004optimizing}
Xue, G.R., Zeng, H.J., Chen, Z., Yu, Y., Ma, W.Y., Xi, W., Fan, W.: Optimizing
  web search using web click-through data. In: Proceedings of the thirteenth
  ACM international conference on Information and knowledge management. pp.
  118--126 (2004)

\bibitem{zamani2020generating}
Zamani, H., Dumais, S., Craswell, N., Bennett, P., Lueck, G.: Generating
  clarifying questions for information retrieval. In: Proceedings of The Web
  Conference 2020. pp. 418--428 (2020)

\bibitem{zamani2020mimics}
Zamani, H., Lueck, G., Chen, E., Quispe, R., Luu, F., Craswell, N.: Mimics: A
  large-scale data collection for search clarification. In: Proceedings of the
  29th ACM International Conference on Information \& Knowledge Management. pp.
  3189--3196 (2020)

\bibitem{zamani2020analyzing}
Zamani, H., Mitra, B., Chen, E., Lueck, G., Diaz, F., Bennett, P.N., Craswell,
  N., Dumais, S.T.: Analyzing and learning from user interactions for search
  clarification. arXiv preprint arXiv:2006.00166  (2020)

\end{thebibliography}
%


\end{document}